\documentclass[review]{elsarticle}

\makeatletter
\def\ps@pprintTitle{%
 \let\@oddhead\@empty
 \let\@evenhead\@empty
 \def\@oddfoot{}%
 \let\@evenfoot\@oddfoot}
\makeatother

\usepackage{lineno,hyperref}
\modulolinenumbers[5]

\begin{document}

\begin{frontmatter}

\title{Faster Deep Ensemble Averaging for Quantification of DNA Damage from Comet Assay Images With Uncertainty Estimates}

\author[mymainaddress]{Srikanth Namuduri\corref{mycorrespondingauthor}}
\author[mymainaddress]{Prateek Mehta}
\author[mysecondaryaddress]{Lise Barbe}
\author[mysecondaryaddress]{Stephanie Lam}
\author[mysecondaryaddress]{Zohreh Faghihmonzavi}
\author[mysecondaryaddress]{Steve Finkbeiner}

\author[mymainaddress]{Shekhar Bhansali}
\cortext[mycorrespondingauthor]{Corresponding author}

\address[mymainaddress]{Electrical \& Computer Engineering, Florida International University, Miami, FL 33174}
\address[mysecondaryaddress]{Finkbeiner Lab, Gladstone Institutes, San Francisco, CA 94158}

\begin{abstract}
Several neurodegenerative diseases involve the accumulation of cellular DNA damage. Comet assays are a popular way of estimating the extent of DNA damage. Current literature on the use of deep learning to quantify DNA damage presents an empirical approach to hyper-parameter optimization and does not include uncertainty estimates. Deep ensemble averaging is a standard approach to estimating uncertainty but it requires several iterations of network training, which makes it time-consuming. Here we present an approach to quantify the extent of DNA damage that combines deep learning with a rigorous and comprehensive method to optimize the hyper-parameters with the help of statistical tests. We also use an architecture that allows for a faster computation of deep ensemble averaging and performs statistical tests applicable to networks using transfer learning. We applied our approach to a comet assay dataset with more than 1300 images and achieved an $R^2$ of 0.84, where the output included the confidence interval for each prediction. The proposed architecture is an improvement over the current approaches since it speeds up the uncertainty estimation by 30X while being statistically more rigorous.

\end{abstract}


\end{frontmatter}

\section{Introduction}
Diseases such as Alzheimer’s, Parkinson’s, ALS, frontotemporal body dementia, and Huntington’s involve a gradual deterioration of the central nervous system and are called neurodegenerative diseases.   Such deterioration is associated with DNA damage, including breaks to the DNA helix.  

DNA damage pathways are altered in neurodegenerative diseases \cite{bradley2014nucleic, madabhushi2014dna, guerrero2016tdp, lopez2016poly, farg2017dna, milanese2018activation,shanbhag2019early}, and DNA double-strand breaks (DSBs) increase with the extent of damage in diseased neurons \cite{enokido2010mutant, jeon2012deregulation, ferlazzo2014mutations, castaldo2019dna, illuzzi2009dna}. This observation suggests a role for DNA damage in neuronal health and indicates that pathways that control it could be targeted to reduce neuronal degeneration.
 Comet assays are used to assess DNA damage in single cells.  In a comet assay, single cells are immobilized on a slide in an agarose gel, lysed in place, and subjected to an electric field. Breaks in the DNA allow it to migrate farther in the electric field than intact DNA can.  The migrating DNA forms a trail that looks like the tail of a comet; the bulk of slow-moving, undamaged DNA forms the comet’s head. The percentage of DNA material in the tail of the comet is a measure of the extent of DNA damage.  Damaged DNA is a marker for genotoxicity, and hence, comet assays can be used for screening drugs \cite{karbaschi2014novel}\cite{koppen2017next}.

Calculating the extent of damage from the comet assay images requires an image-processing tool to estimate the number of pixels and pixel intensity values in the head and the tail. In the past, this was done using classical image processing techniques \cite{lee2018hicomet}, but recently machine learning (ML) has been leveraged by researchers to automate the scoring process \cite{namuduri2019automated}.

Among the various ML techniques available, deep learning (DL) approaches, specifically Convolutional Neural Networks (CNNs), do not require manual feature engineering and are the most effective when working with images. When deep learning is applied to regression problems, i.e., the target variable is continuous, it is called deep regression. 

When using ML or DL to assess the extent of damage in human cells, such as in neurological diseases, accuracy and rigor are both critical. Otherwise, the results cannot be trusted or relied upon. In the existing literature, the emphasis is on accuracy and not reliability. There is a need to improve the rigorousness of the deep learning approaches for better acceptance in fields such as medicine. The goal of the research presented in this paper was to identify the deep regression approach that can quantify the percentage of DNA damage from comet assay images in a rigorous fashion.

Any estimate made using regression techniques involves some uncertainty. In problems such as assessing the DNA damage from images, it is essential to quantify the uncertainty because it conveys the extent to which the actual value may differ from the estimated value. The user then decides whether the reported uncertainty is acceptable. 

One of the challenges with the application of DL in medicine and in neurodegenerative diseases research is insufficient data. It cannot be expected that millions of comet assay images from diseased cells will be available for training. Even with limited data, the other challenge is the time taken for computation. Transfer learning and an application of multi-task learning to uncertainty estimation were explored in this research to address these challenges. The methods presented in this paper can be applied in any domain where deep regression is being used. The architecture presented in this paper can be used to improve the reliability of deep regression without sacrificing the speed of computation.

\subsection{Deep Learning}

Images do not conform to a tabular structure and are considered un-structured data. They consist of a matrix of pixel values and can be treated as two-dimensional arrays for computational purposes. Deep Learning (DL) approaches are suitable for extracting features and making estimations from such
data. While traditional Machine Learning (ML) approaches can also be applied for unstructured data, DL approaches give superior accuracy. Hence the scope of this paper is restricted to DL.
DL is a subset of ML that uses neural networks with several layers. The feature learning is hierarchical, with each layer learning from the previous one.

The layered architecture allows the DL networks to learn complex relationships between the features and the target variable. In comet assays assessment, the target variable is the percentage of DNA in the tail.

\begin{equation}
\label{function}
y  = f(x;\theta) \,
\end{equation}

\begin{equation}
\label{chain}
f(x)  = f_3(f_2(f_1(x))) \,
\end{equation}

The layers in the network consist of weights, represented by $\theta$ in equation \ref{function}. The learning process involves updating the weights in the layers using the backpropagation algorithm \cite{Lecun2015}, and \cite{goodfellow2016deep}. The learned relationship can be represented as a sequence of nested functions as shown in equation \ref{chain}. 

When a deep learning architecture is used to perform regression, it is called deep regression. The focus of this paper is on deep regression.

\subsection{Convolutional Neural Networks (CNN)}

Convolution is a mathematical operation defined on two functions. In the case of continuous functions, it consists of an integral of the product of one function and a reversed and shifted version of the second function. The discrete convolution operation consists of summation is shown in \ref{convolution}. The discrete version of convolutions is the mathematical basis for the CNNs. 

\begin{figure}[h!]
 \centering
    \includegraphics[scale=0.5]{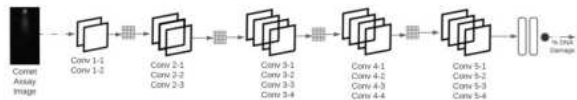}
    \caption{A VGG19 network with 2 fully connected layers. The input is a Comet Assay image and the output is the \% DNA damage}
    \label{fig:CNN}
\end{figure}

A neural network layer that uses the convolution operation is a convolutional layer. Figure \ref{fig:CNN} shows a representative image of a CNN applied to an image. A CNN consists of a convolutional block, and the fully connected block \cite{Lecun2015, namuduri2019automated}.

The discrete convolution operation, shown in equation \ref{convolution}, is the mathematical basis for the convolutional layers. $I$ indicates the input, and $K$ denotes a filter applied to the input. The output of each convolutional layer indicated by $C$ in Equation \ref{convolution} is referred to as the feature map \cite{goodfellow2016deep}. 

\begin{equation}
    C(i,j) = \sum_m \sum_n I(m,n)K(i-m,j-n)
    \label{convolution}
\end{equation}

\subsection{Transfer Learning}
Transfer learning involves a machine learning model leveraging the learning of another model. The transfer of knowledge can be from a related learning task or a completely unrelated task. It leads to faster learning and better performance on the second task in both cases. Another major advantage is that the second task can be trained with a smaller dataset. \cite{weiss2016survey, bengio2012deep, cook2013transfer, tan2017distant, pan2009survey,tajbakhsh2016convolutional}

In the case of DL, transfer of knowledge is achieved by freezing (or not training) some of the network layers after training on the first task and then training or fine-tuning the rest of the layers to learn the second task. Typically the initial layers are kept, and the subsequent layers are fine-tuned for the second or final task.\cite{tajbakhsh2016convolutional} 

The initial training (first task) can be done from scratch by selecting a task with sufficient data. Alternatively, a pre-trained model can be used. Trained weights for some of the popular DL architectures like VGG and Resnet, on well-known datasets such as MNIST and ImageNet, have been open-sourced by the respective researchers to be used for transfer learning. The study reported here used such a VGG19 network, pre-trained on ImageNet data.

\subsection{Multi-Task Learning}
The default approach trains a neural network to learn a single task based on a specific metric. However, by only focusing on a single task, the network may ignore other information that may be useful. When there are several related tasks, the respective networks can gain better information from each other’s learning by sharing representations between the tasks. With comet assays, the head diameter, the tail length, and the number of pixels in the head are all related variables. They are all different aspects of the comet but are related to determining the amount of DNA damage.

Multi-Task Learning (MTL) involves using networks with shared layers to learn multiple tasks together, enabling the models to generalize better on the original task. This also leads to several benefits, such as less computational memory, faster training in some cases, and better generalization of learning due to regularization.
\cite{zhang2021survey}.

\section{Related Work}
Researchers have applied Deep Learning (DL), specifically Convolutional Neural Networks (CNNs), to comet assay images. However, most applications of DL to comet assays focus on classifying the images. Only a few studies have attempted to quantify the extent of damage from the images, and none have attempted to calculate a confidence interval for their estimates. Deep learning is a stochastic process, and every estimate involves some uncertainty, and not reporting the uncertainty can mislead the user. 

Studies that analyzed comet images via DL have successfully applied transfer learning \cite{zhang2021survey}. The research presented in this paper also leverages transfer learning. Multi-task learning (MTL) has been applied to DL by several researchers.  The neural network presented in this paper involves a novel use of the MTL architecture to calculate the confidence intervals. \cite{anarossi2019comet}  \cite{rosati2020faster}

\section{Data Collection and Annotation}
The Trevigen comet assay kit (4250-050-ESK, R\&D Systems) was used for running the experiments. Some cells were processed using the alkaline comet assay protocol, and the others using the neutral comet assay protocol. The slides used during the experiments were imaged with fluorescent microscopy. Specifically, a Nikon Ti-E inverted microscope was used with a 10x/0.45 objective.

Images from both neutral and alkaline protocols can train the deep learning model. A total of 1309 images were collected from the experiments. Out of these images, 1047 were used for training, and 262 were used to test the model.

\begin{figure}[htbp]
\centering
\includegraphics[width=\linewidth]{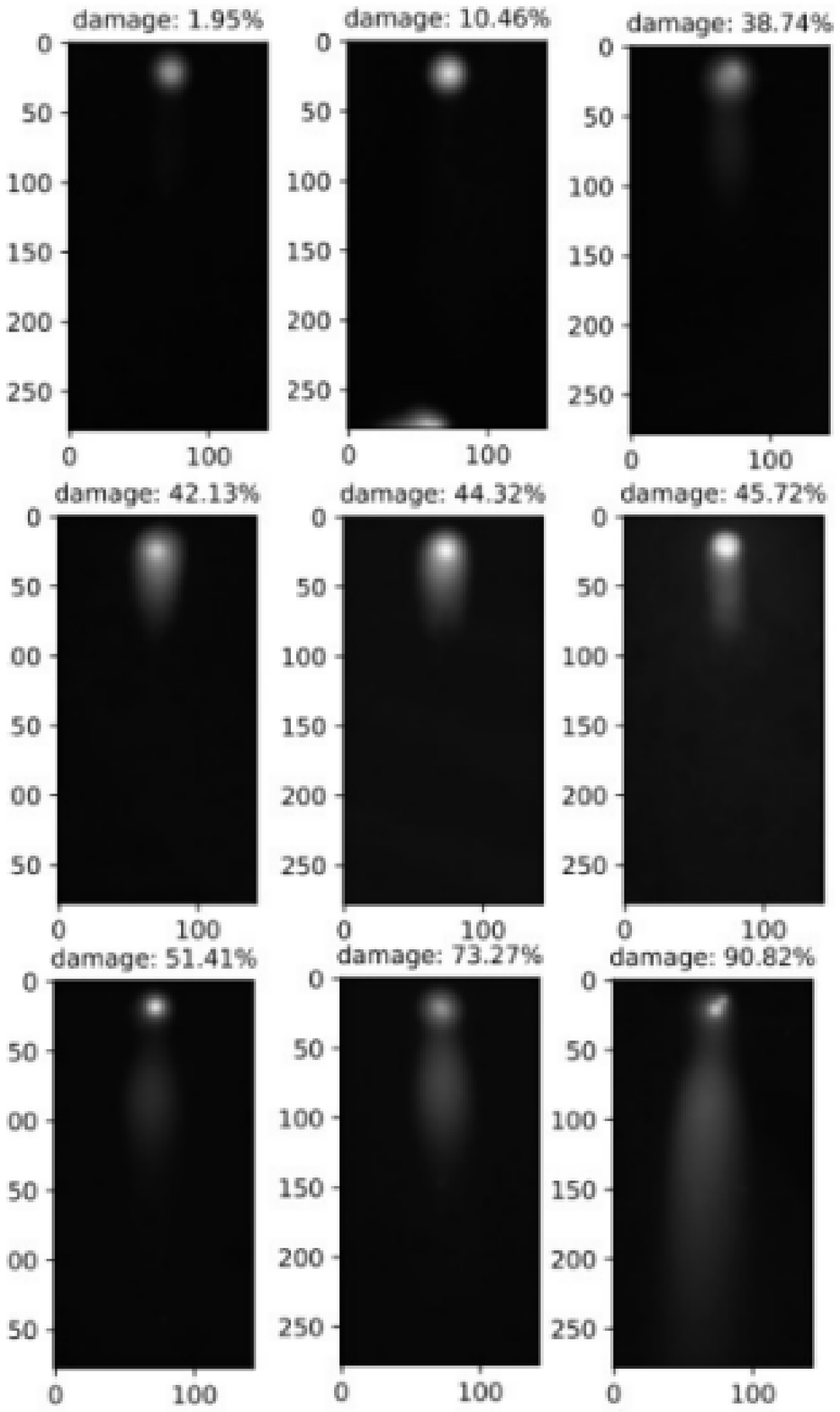}
\caption{Images of some of the comets in the training data. The comet shape is not always apparent to the naked eye}
\label{fig:comet_viz}
\end{figure}

\subsection{Comet Assay Scoring}
The ground truth was generated with an algorithm that detects, segments, and quantifies the comets in the microscopy images. In short, we start with a re-orientation of the comets as needed to have the tails facing downward of the heads. Then, binary thresholding is performed to separate meaningful objects from the background. Noise and comets out of bound were removed, while the rest of the detected comet objects were cropped into 273 x 143-pixel size (with 0.6465 µM per pixel). Any crop containing only tail objects, overlapping comets, or artifacts is filtered using size, circularity, and intensity criteria. During segmentation, comet heads and tails are identified using different thresholding. Different measurements are then calculated to quantify DNA damage, including head diameter, tail length, area, DNA content, and length. All values are saved into a CSV for any further quantitative analysis.

\section{Methodology}
A 2019 paper \cite{lathuiliere2019comprehensive} has noted that most of the research presented in the current literature on deep learning lacks a statistical evaluation of the performance. A lack of confidence interval reporting was also noted in the context of deep regression. The paper then presents a rigorous approach to optimizing a deep learning network. This paper applies the same approach to optimizing the network when applied to the comet assay data, but with more samples per configuration.

\cite{lathuiliere2019comprehensive} demonstrated that, for Deep Regression, using a standard architecture such as VGG16 or Resnet and rigorously optimizing the hyper-parameters gives the same or better performance as compared to more complex architectures. This paper leverages this finding by using a VGG19 architecture and rigorously optimizing the parameters.

\cite{jimenez2020drug} notes that uncertainty is ubiquitous in natural sciences, but DL methods in current use do not pay attention to uncertainty estimation. One of the sections in the paper captures several methods to estimate the uncertainty and mentions that Deep Ensemble Averaging is the most popular approach.

The approach presented in this paper is a quicker way to perform Deep Ensemble Averaging. The ensemble of networks is trained in parallel rather than serial to hasten the process. It is a trade-off between computation time and memory. However, the use of additional memory is acceptable (in most cases), given the speed up. 

Further, the Multi-Task Learning architecture presented also reduces the compute time for the statistical tests. Multiple training runs of the network (to get enough samples to perform a statistical test) are replaced by a single run (that gives enough samples).

\subsection{Metrics}
The neural network's output is a continuous variable, namely the percentage damage. Hence, it is treated as a regression problem. Since the training uses annotated or labeled samples, it is supervised learning. The cost function optimized during the learning process is the mean square error.  

\begin{equation}
\label{mse}
MSE = \frac{1}{n}\sum_{i=1}^{n} {(\hat{y}-y_i)}^{2} \,
\end{equation}

The coefficient of determination ( R\^2) indicates the amount of variation in the data explained by the model. R2 is defined as follows:  

\begin{equation}
\label{R2}
R^2 = 1 - \frac{SSE}{SST} \,
\end{equation}

In equation \ref{R2}, SSE is the Sum of Squares Error of the model, and SST is the Sum of Squares Total. SST is an estimate of the total variation in the data.

\subsection{Data Preprocessing}
The dataset used for this research included only grayscale images. Any out-of-sample images were converted to grayscale before being included in our model. All the images were re-sized to be of the same size for training and testing purposes.
Data augmentation techniques were applied to improve the model robustness and increase the size of the training dataset. This involved image rotation, zoom-in/zoom-out (limited magnitude, so as to not cut off the comets) , flipping, and translation operations. These operations modify the images without changing their aspect ratios and hence do not impact the ratio of the tail size to the head size.

\subsection{Computational Environment}
The pre-processing, training, and testing were done on Google Colab computational environment. The code was written in Python 3, and the packages that were indispensable for this implementation were NumPy, Pandas, and Keras. The plotting was done using the Matplotlib library in python. The network layers table/visualization was generated using the build-in plot-model facility from Keras. 

\subsection{Network Architecture}
Convolutional Neural Networks (CNNs) consists of one or more convolutional layers and fully connected layers. Deeper networks, which have more layers, can learn more complex tasks. VGG19 \cite{simonyan2014very} networks use a total of 19 layers that contain weights, which includes five blocks of convolutional layers interspersed with pooling layers and fully connected layers. 

Prior research from this group found that using a pre-trained VGG19 network performed better than other models\cite{namuduri2019automated}. As an extension, the same pre-trained convolutional blocks from VGG19 are used, but the fully connected block uses an MTL type architecture, where 30 replications of the dense layers are connected in parallel. Since the weights of the dense layers are randomly initialized, the learned weights at the end of the training process are also different. Consequently, the estimates made by each of the 30 branches are slightly different. 

\begin{figure}[htbp]
\centering
\includegraphics[width=\linewidth]{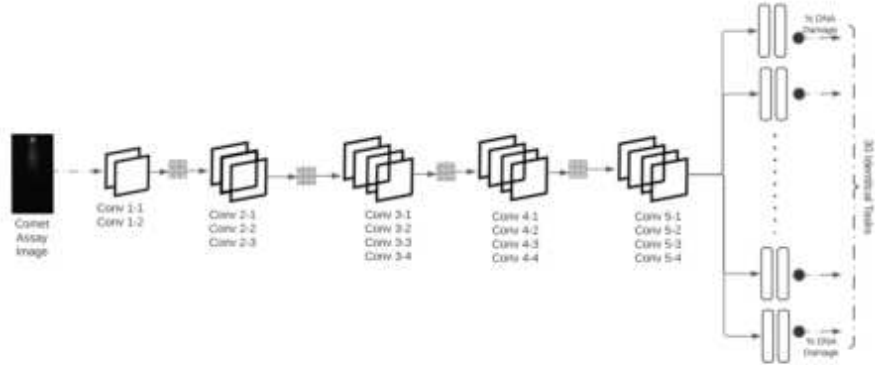}
\caption{A Multi-task learning architecture, with common convolutional layers and separate dense layers, re-purposed to learn the same task 30 times. Each of the outputs is an estimate of the percentage DNA damage }
\label{fig:mlt_ci}
\end{figure}

The VGG19 network and the weights pre-trained on the Imagenet dataset were loaded from the Keras library, like in \cite{namuduri2019automated}. The convolutional layers were frozen and unchanged during training, but the dense layers went through learning. Linear activation is used in the final layer since the output is a continuous variable.

This architecture was implemented using the Keras Functional API, which allows the creation of deep learning models with shared layers. The Keras Functional API developer guide can be found at \cite{web:keras_functional}. The examples in the developer guide helped develop the code for this research. The output from the convolutional block was used as input for 30 different fully connected layers, which led to 30 outputs. A simple for-loop was used to generate these 30 layers.

\subsection{Hyper-parameter Optimization}
The optimization approach used in this research is based on the work of \cite{lathuiliere2019comprehensive}. Where \cite{lathuiliere2019comprehensive} used five samples to run statistical tests for each of the configurations, this research used a parallel architecture to obtain 30 samples for the same.

Optimizing a DL network requires finding the optimal combination of hyper-parameters, such as the number of nodes in each layer, learning rate, and batch size. DL models are inherently stochastic, with the weight initialization, optimizer, and the samples in each batch being some of the sources of randomness. Due to this, training the network repeatedly with the same set of parameters results in slightly different estimations of the DNA damage percentage. So statistical tests are required to compare one configuration of hyper-parameters with another. 

In all the comet assay papers in prior literature and most of the papers on deep regression, an empirical approach is taken in finding the right combination of hyper-parameters. In contrast, training each configuration multiple times, followed by statistical tests for comparisons, the best set of hyper-parameters can be picked based on statistical significance rather than the empirical results.  \cite{lathuiliere2019comprehensive}. 

This research uses the Wilcoxon Signed Rank test to compare each combination, like in \cite{lathuiliere2019comprehensive}. Even though the number of samples used in this research is higher, we stick to this test and do not use a parametric test since there is no basis for assuming that the underlying distribution is Gaussian. 

First, a list of hyper-parameter combinations is generated, and a baseline combination is chosen. The rest of the combinations are compared with the baseline until we find a better performing combination, at which point, the baseline is updated. After this, the subsequent combinations are compared to the new baseline until an even better one is found. Comparing and finding a better baseline is repeated until all the combinations are exhausted. The median R\^2 of the best combination is considered the coefficient of determination of the optimized model.

While comparing a configuration with the baseline, the null hypothesis is as follows:

"\textit{$H_0$: The $R^2$ of the current combination is not significantly better than the baseline}"

Whenever the P-value is less than 0.05, we reject the null hypothesis, which means the new combination is better than the baseline. This method of optimizing the model is more scientific and hence more reliable and trustworthy than an empirical approach.

\subsection{Uncertainty Estimation}
 
The $R^2$ of a model alone does not convey the trustworthiness of the estimates. Reporting the confidence interval (CI) can take us further in understanding the estimates. CI indicates the reliability of the process involved in the estimation. It is a range of values where the actual value can be found with a certain confidence level. The most common way to estimate CI in DL methods is the Deep Ensemble Averaging (DEA)  \cite{jimenez2020drug}. 

As the name suggests, DEA involves creating an ensemble of predictors to calculate the predicted value and the CI. In typical implementations of DEA, the same model is trained several times (typically 30), and then the CI is calculated as:

\begin{equation}
\label{CI}
CI = (\mu - \sigma, \mu + \sigma) \,
\end{equation}

In equation \ref{CI}, $\mu$ is the mean of the 30 estimates and $\sigma$ is the standard deviation. 

The predicted value is calculated as:
\begin{equation}
\label{pred}
R^2 = \frac{\Sigma(R_i^2)}{n} \,
\end{equation}

In equation \ref{pred}, $R_i$ is the $i^th$ sample and $n$ is the total number of samples.

Even though DEA is a well-known approach, running a model several times is very time-consuming. When using large datasets (with millions of images), each training run can take hours to days, depending on the computing infrastructure available. 

Being able to run several instances of the same network in parallel can speed up the process of using statistical tests to optimize the network and the calculation of CI. The network architecture proposed in this paper serves that purpose.  CI can be calculated by replicating the same task in an MTL type configuration to get 30 different predictions. The predictions are different because of the underlying stochastic process.

The architecture suggested here is best suited when using transfer learning, where we freeze the initial convolutional layers and fine-tune the fully connected layers.  This way, doing a single training run with the new architecture is equivalent to doing 30 runs of the network with a single output.

\section{Results}
The performance of a VGG19 was compared to that of an MTL-VGG19 network trained on the same set of comet images. The VGG19 network was trained 30 times, making predictions on the test set each time. Subsequently, CI and the predicted value were calculated for each test sample. In contrast, the MTL-VGG19 only needed training once and output 30 predictions for each test sample. CI and the predicted value were calculated for the MTL-VGG19 network.

The average size of the CI for the test samples using the VGG19 network (and 30 training runs) was 1.89 and for the MTL-VGG19 architecture was 1.36.     The following was the null hypothesis to compare the two groups:

"\textit{$H_0$: There is no significant difference in the CI calculated using VGG19 network and the CI calculated using MLT-VGG19}"
The p-value when performing a Student’s T-test was 0.40, which is not significant. So we could not reject the null hypothesis. Hence the CI obtained using both approaches can be concluded to be the same.  Hence the MTL-VGG19 architecture can calculate the CI in place of the existing method, which is more time-consuming. 

Once it was established that the results from the new architecture are not different from repeated training of the original network, it was used to optimize the network and calculate the median R\^2 and CI. The results (not exhaustive) from optimizing the network are shown in table \ref{tab:perf}. The first five columns of the table show the various hyper-parameters being optimized. The last two columns show the Test $R^2$ and the P-value, respectively. The P-value in each row results from the statistical test conducted to compare the configuration with the one in the previous row. If the Test $R^2$ over 30 iterations is significantly different from the previous configuration, the P-value is expected to be less than 0.05. The first row shows the baseline configuration. Only one of the other configurations was significantly better than the baseline, as shown in row 4, indicating the final optimized set of hyper-parameters.

\begin{table}[h!]
\label{tab:perf}
\begin{center}       
\begin{tabular}{|l|l|l|l|l|l|l|}
\hline
\rule[-1ex]{0pt}{3.5ex}  Nodes1 & Nodes2 & Drop Out  & LR &  Batch Size & Test R$^2$ & 
P-Value \\
\hline

\rule[-1ex]{0pt}{3.5ex}  256 & 128 & 0.2 & 0.01 & 32& 0.69 & NA*  \\
\hline
\rule[-1ex]{0pt}{3.5ex}  256 & 128 & 0.2 & 0.01 & 64& 0.73 & 0.99   \\
\hline
\rule[-1ex]{0pt}{3.5ex}  256 & 128 & 0.2 & 0.001 & 16& 0.76 & 0.35   \\
\hline

\rule[-1ex]{0pt}{3.5ex}  256 & 128 & 0.2 & 0.001 & 32& 0.84 & \(8.67 \times 10^-7\) **  \\
\hline
\rule[-1ex]{0pt}{3.5ex}  256 & 128 & 0.2 & 0.001 & 64& 0.82 & 0.99 \\
\hline
\rule[-1ex]{0pt}{3.5ex}  256 & 128 & 0.2 & 0.005 & 16& 0.79 & 0.99    \\
\hline

\rule[-1ex]{0pt}{3.5ex}  256 & 128 & 0.2 & 0.005 & 32& 0.81 & 0.99    \\
\hline
\rule[-1ex]{0pt}{3.5ex}  256 & 128 & 0.4 & 0.01 & 16& 0.57 & 0.99    \\
\hline
\rule[-1ex]{0pt}{3.5ex}  256 & 128 & 0.4 & 0.001 & 16& 0.71 & 0.99 \\
\hline

\rule[-1ex]{0pt}{3.5ex}  128 & 64 & 0.2 &  0.01& 16 & 0.70 & 0.99 \\
\hline
\rule[-1ex]{0pt}{3.5ex}  128 & 64 & 0.2 &  0.01& 32 & 0.74 & 0.99  \\
\hline
\rule[-1ex]{0pt}{3.5ex}  128 & 64 & 0.2 &  0.01& 64 & 0.78 & 0.99   \\
\hline

\rule[-1ex]{0pt}{3.5ex}  128 & 64 & 0.2 &  0.001& 16 & 0.83 & 0.99 \\
\hline
\rule[-1ex]{0pt}{3.5ex}  128 & 64 & 0.2 &  0.001& 32 & 0.83 & 0.99  \\
\hline
\rule[-1ex]{0pt}{3.5ex}  128 & 64 & 0.2 &  0.001& 64 & 0.82 & 0.99   \\
\hline

\rule[-1ex]{0pt}{3.5ex}  128 & 64 & 0.2 &  0.005& 32 & 0.79 & 0.99 \\
\hline
\rule[-1ex]{0pt}{3.5ex}  128 & 64 & 0.4 &  0.01& 16 & 0.39 & 0.99  \\
\hline
\rule[-1ex]{0pt}{3.5ex}  128 & 64 & 0.4 &  0.001& 16 & 0.70 & 0.99   \\
\hline

\end{tabular}
\caption{Performance of various Hyper-Parameter Combinations. * Baseline Configuration. ** Significant. LR - Learning Rate} 
\end{center}
\end{table}

\section{Conclusion and Future Work}

Using the proposed architecture, in conjunction with transfer learning, we calculated confidence intervals in one shot, i.e., a single training run. This architecture can be used in most deep regression problems, and it offers a way to calculate CI and perform statistical tests much faster. The speed could be increased up to 30-fold with 30 samples, which is typical for calculating CI and statistical tests.

So far in this research, all the measurements available from the comet images have not been leveraged. Learning multiple measurements such as the diameter of the head, length of the tail can improve the overall accuracy and reliability of the damage assessment. Our research group plans to study the use of multiple measurements and assess the impact on accuracy to further the research in the use of Deep Learning for scoring the comets while studying neurodegenerative diseases.

\section{Acknowledgements}

This work was made possible with support from NIH grants R01 LM013617, R37 NS101996, P01 AG054407 and RF1 AG058476 to S.F. as well as support from the Koret Foundation Artificial Intelligence Program for Biomedical Research

\bibliography{mybibfile}

\begin{thebibliography}{10}
\expandafter\ifx\csname url\endcsname\relax
  \def\url#1{\texttt{#1}}\fi
\expandafter\ifx\csname urlprefix\endcsname\relax\def\urlprefix{URL }\fi
\expandafter\ifx\csname href\endcsname\relax
  \def\href#1#2{#2} \def\path#1{#1}\fi

\bibitem{bradley2014nucleic}
M.~A. Bradley-Whitman, M.~D. Timmons, T.~L. Beckett, M.~P. Murphy, B.~C. Lynn,
  M.~A. Lovell, Nucleic acid oxidation: an early feature of alzheimer's
  disease, Journal of neurochemistry 128~(2) (2014) 294--304.

\bibitem{madabhushi2014dna}
R.~Madabhushi, L.~Pan, L.-H. Tsai, Dna damage and its links to
  neurodegeneration, Neuron 83~(2) (2014) 266--282.

\bibitem{guerrero2016tdp}
E.~N. Guerrero, H.~Wang, J.~Mitra, P.~M. Hegde, S.~E. Stowell, N.~F. Liachko,
  B.~C. Kraemer, R.~M. Garruto, K.~Rao, M.~L. Hegde, Tdp-43/fus in motor neuron
  disease: complexity and challenges, Progress in neurobiology 145 (2016)
  78--97.

\bibitem{lopez2016poly}
R.~Lopez-Gonzalez, Y.~Lu, T.~F. Gendron, A.~Karydas, H.~Tran, D.~Yang,
  L.~Petrucelli, B.~L. Miller, S.~Almeida, F.-B. Gao, Poly (gr) in
  c9orf72-related als/ftd compromises mitochondrial function and increases
  oxidative stress and dna damage in ipsc-derived motor neurons, Neuron 92~(2)
  (2016) 383--391.

\bibitem{farg2017dna}
M.~A. Farg, A.~Konopka, K.~Y. Soo, D.~Ito, J.~D. Atkin, The dna damage response
  (ddr) is induced by the c9orf72 repeat expansion in amyotrophic lateral
  sclerosis, Human molecular genetics 26~(15) (2017) 2882--2896.

\bibitem{milanese2018activation}
C.~Milanese, S.~Cerri, A.~Ulusoy, S.~V. Gornati, A.~Plat, S.~Gabriels,
  F.~Blandini, D.~A. Di~Monte, J.~H. Hoeijmakers, P.~G. Mastroberardino,
  Activation of the dna damage response in vivo in synucleinopathy models of
  parkinson’s disease, Cell death \& disease 9~(8) (2018) 1--12.

\bibitem{shanbhag2019early}
N.~M. Shanbhag, M.~D. Evans, W.~Mao, A.~L. Nana, W.~W. Seeley, A.~Adame, R.~A.
  Rissman, E.~Masliah, L.~Mucke, Early neuronal accumulation of dna double
  strand breaks in alzheimer’s disease, Acta neuropathologica communications
  7~(1) (2019) 1--18.

\bibitem{enokido2010mutant}
Y.~Enokido, T.~Tamura, H.~Ito, A.~Arumughan, A.~Komuro, H.~Shiwaku, M.~Sone,
  R.~Foulle, H.~Sawada, H.~Ishiguro, et~al., Mutant huntingtin impairs
  ku70-mediated dna repair, Journal of Cell Biology 189~(3) (2010) 425--443.

\bibitem{jeon2012deregulation}
G.~S. Jeon, K.~Y. Kim, Y.~J. Hwang, M.-K. Jung, S.~An, M.~Ouchi, T.~Ouchi,
  N.~Kowall, J.~Lee, H.~Ryu, Deregulation of brca1 leads to impaired
  spatiotemporal dynamics of $\gamma$-h2ax and dna damage responses in
  huntington’s disease, Molecular neurobiology 45~(3) (2012) 550--563.

\bibitem{ferlazzo2014mutations}
M.~L. Ferlazzo, L.~Sonzogni, A.~Granzotto, L.~Bodgi, O.~Lartin, C.~Devic,
  G.~Vogin, S.~Pereira, N.~Foray, Mutations of the huntington’s disease
  protein impact on the atm-dependent signaling and repair pathways of the
  radiation-induced dna double-strand breaks: Corrective effect of statins and
  bisphosphonates, Molecular neurobiology 49~(3) (2014) 1200--1211.

\bibitem{castaldo2019dna}
I.~Castaldo, M.~De~Rosa, A.~Romano, C.~Zuchegna, F.~Squitieri, R.~Mechelli,
  S.~Peluso, C.~Borrelli, A.~Del~Mondo, E.~Salvatore, et~al., Dna damage
  signatures in peripheral blood cells as biomarkers in prodromal huntington
  disease, Annals of neurology 85~(2) (2019) 296--301.

\bibitem{illuzzi2009dna}
J.~Illuzzi, S.~Yerkes, H.~Parekh-Olmedo, E.~B. Kmiec, Dna breakage and
  induction of dna damage response proteins precede the appearance of visible
  mutant huntingtin aggregates, Journal of neuroscience research 87~(3) (2009)
  733--747.

\bibitem{karbaschi2014novel}
M.~Karbaschi, M.~S. Cooke, Novel method for the high-throughput processing of
  slides for the comet assay, Scientific reports 4~(1) (2014) 1--6.

\bibitem{koppen2017next}
G.~Koppen, A.~Azqueta, B.~Pourrut, G.~Brunborg, A.~R. Collins, S.~A. Langie,
  The next three decades of the comet assay: a report of the 11th international
  comet assay workshop, Mutagenesis 32~(3) (2017) 397--408.

\bibitem{lee2018hicomet}
T.~Lee, S.~Lee, W.~Y. Sim, Y.~M. Jung, S.~Han, J.-H. Won, H.~Min, S.~Yoon,
  Hicomet: a high-throughput comet analysis tool for large-scale dna damage
  assessment, BMC bioinformatics 19~(1) (2018) 49--61.

\bibitem{namuduri2019automated}
S.~Namuduri, B.~N. Narayanan, M.~Karbaschi, M.~Cooke, S.~Bhansali, Automated
  quantification of dna damage via deep transfer learning based analysis of
  comet assay images, in: Applications of Machine Learning, Vol. 11139,
  International Society for Optics and Photonics, 2019, p. 111390Y.

\bibitem{Lecun2015}
Y.~Lecun, Y.~Bengio, G.~Hinton, {Deep learning} (may 2015).
\newblock \href {https://doi.org/10.1038/nature14539}
  {\path{doi:10.1038/nature14539}}.

\bibitem{goodfellow2016deep}
I.~Goodfellow, Y.~Bengio, A.~Courville, Deep learning, MIT press, 2016.

\bibitem{weiss2016survey}
K.~Weiss, T.~M. Khoshgoftaar, D.~Wang, A survey of transfer learning, Journal
  of Big data 3~(1) (2016) 1--40.

\bibitem{bengio2012deep}
Y.~Bengio, Deep learning of representations for unsupervised and transfer
  learning, in: Proceedings of ICML workshop on unsupervised and transfer
  learning, JMLR Workshop and Conference Proceedings, 2012, pp. 17--36.

\bibitem{cook2013transfer}
D.~Cook, K.~D. Feuz, N.~C. Krishnan, Transfer learning for activity
  recognition: A survey, Knowledge and information systems 36~(3) (2013)
  537--556.

\bibitem{tan2017distant}
B.~Tan, Y.~Zhang, S.~Pan, Q.~Yang, Distant domain transfer learning, in:
  Proceedings of the AAAI Conference on Artificial Intelligence, Vol.~31, 2017.

\bibitem{pan2009survey}
S.~J. Pan, Q.~Yang, A survey on transfer learning, IEEE Transactions on
  knowledge and data engineering 22~(10) (2009) 1345--1359.

\bibitem{tajbakhsh2016convolutional}
N.~Tajbakhsh, J.~Y. Shin, S.~R. Gurudu, R.~T. Hurst, C.~B. Kendall, M.~B.
  Gotway, J.~Liang, Convolutional neural networks for medical image analysis:
  Full training or fine tuning?, IEEE transactions on medical imaging 35~(5)
  (2016) 1299--1312.

\bibitem{zhang2021survey}
Y.~Zhang, Q.~Yang, A survey on multi-task learning, IEEE Transactions on
  Knowledge and Data Engineering (2021).

\bibitem{anarossi2019comet}
E.~Anarossi, R.~D. Yanuaryska, F.~U. Nuha, S.~Mulyana, et~al., Comet assay
  classification for buccal mucosa’s dna damage measurement with super tiny
  dataset using transfer learning, in: Asian Conference on Intelligent
  Information and Database Systems, Springer, 2019, pp. 279--289.

\bibitem{rosati2020faster}
R.~Rosati, L.~Romeo, S.~Silvestri, F.~Marcheggiani, L.~Tiano, E.~Frontoni,
  Faster r-cnn approach for detection and quantification of dna damage in comet
  assay images, Computers in Biology and Medicine 123 (2020) 103912.

\bibitem{lathuiliere2019comprehensive}
S.~Lathuili{\`e}re, P.~Mesejo, X.~Alameda-Pineda, R.~Horaud, A comprehensive
  analysis of deep regression, IEEE transactions on pattern analysis and
  machine intelligence 42~(9) (2019) 2065--2081.

\bibitem{jimenez2020drug}
J.~Jim{\'e}nez-Luna, F.~Grisoni, G.~Schneider, Drug discovery with explainable
  artificial intelligence, Nature Machine Intelligence 2~(10) (2020) 573--584.

\bibitem{simonyan2014very}
K.~Simonyan, A.~Zisserman, Very deep convolutional networks for large-scale
  image recognition, arXiv preprint arXiv:1409.1556 (2014).

\bibitem{web:keras_functional}
F.~Chollet, \href{https://keras.io/guides/functional_api/}{Keras functional
  api}, last accessed December 6th 2021 (2019).
\newline\urlprefix\url{https://keras.io/guides/functional_api/}

\end{thebibliography}

\end{document}